\pdfoutput=1
\documentclass[prl,twocolumn,showpacs,groupedaddress,superscriptaddress,nofootinbib,floatfix,preprintnumbers,longbibliography]{revtex4-1}
\usepackage{amssymb,amsmath,graphicx,color}
\usepackage{physics}
\usepackage{bm}
\usepackage{appendix}
\usepackage[tight]{subfigure}
\usepackage[export]{adjustbox}
\usepackage{braket}
\usepackage[colorlinks=true,citecolor=blue,linkcolor=red]{hyperref}
\usepackage{cleveref}
\usepackage{multirow}
\usepackage{rotating,booktabs}
\usepackage[verbose]{placeins}
\usepackage{color}
\newcommand\bea{\begin{eqnarray}}
\newcommand\eea{\end{eqnarray}}

\begin{document}
\title{
Protecting local and global symmetries in simulating 1+1-D 
non-abelian gauge theories}
\author{Emil Mathew}
\email{p20210036@goa.bits-pilani.ac.in}
\author{Indrakshi Raychowdhury}
\email{indrakshir@goa.bits-pilani.ac.in}
\affiliation{Department of Physics, BITS-Pilani,
K K Birla Goa Campus, Zuarinagar, Goa 403726, India}
\date{\today}
\begin{abstract}
Efficient quantum simulation protocols for any quantum theories demand efficient protection protocols for its underlying symmetries. This task is nontrivial for gauge theories as it is involves local symmetry/invariance. For non-Abelian gauge theories, protecting all the symmetries generated by a set of mutually non-commuting generators, is particularly difficult. In this letter, a global symmetry-protection protocol is proposed. Using the novel loop-string-hadron formalism of non-Abelian lattice gauge theory, we numerically demonstrate that all of the local symmetries get protected even for large time by this global symmetry protection scheme.  With suitable protection strength, the dynamics of a (1+1)-dimensional SU(2) lattice gauge theory remains confined in the physical Hilbert space of the theory even in presence of explicit local symmetry violating terms in the Hamiltonian that may occur in both analog and digital simulation schemes as an error. The whole scheme holds for SU(3) gauge theory as well.
\end{abstract}
\maketitle
\section{Introduction}
\label{sec:intro}
Gauge theories form the backbone of most of modern physics, ranging from the standard model of particle physics to condensed matter systems \cite{rothe2012lattice, fradkin2013field}. Typical calculations/computations for these theories are done via the path-integral approach that suits both analytical and numerical studies. Lattice regularized version of gauge theories \cite{Wilson:1974sk}  provide an extremely useful platform to perform nonperturbative calculations using Monte-Carlo simulation towards understanding the  strong interaction described by quantum chromodynamics (QCD). Despite its usefulness, the numerical lattice-QCD program faces a roadblock due to sign problem towards certain aspects of computation, such as computing dynamics or exploring the full phase diagram of QCD \cite{wiese2014towards}.  Of late, an interdisciplinary community has ramped up efforts to combine aspects of quantum technology with that of lattice gauge theories to tackle some of these outstanding problems \cite{Banerjee:2012pg,Banerjee:2012xg, Zohar:2012xf,Zohar:2013zla,Zohar:2014qma,Zohar:2015hwa, Zohar:2019ygc, klco2018quantum, klco20202, Davoudi:2019bhy, davoudi2021toward, davoudi2020towards, Raychowdhury:2019iki, raychowdhury2020solving, dasgupta2022cold,https://doi.org/10.48550/arxiv.2204.03381}.

The natural framework to study dynamics of gauge theories without any sign problem is a canonical/Hamiltonian framework. The Hamiltonian description of lattice gauge theories, put forth by Kogut and Susskind \cite{Kogut:1974ag}, provide such a framework. 
However, the exponentially growing dimension of the Hilbert space with lattice size, is not suitable for classical computation, and that is where quantum simulation/computation is expected to be useful. However, one additional complication that accompanies the gauge theory Hamiltonian is the local constraints that generate the gauge transformation. Preserving the gauge invariance of the simulated theory with state of the art quantum hardware is a major challenge in quantum simulating lattice gauge theories.
This article presents a protocol to protect the gauge invariance for two dimensional gauge field theories with a continuous non-Abelian gauge group such as SU(2). Protecting all of the symmetries of the theory  should lead to an efficient quantum simulation for the same \cite{tran2021faster}.

The physical or gauge invariant Hilbert space for a gauge theory is defined to be spanned by basis vectors that are annihilated by each and every local constraints or in general by all the Gauss law constraints. The Hamiltonian being gauge invariant, commutes with these constraints, keeping the dynamics confined in the physical Hilbert space.  

Construction of the gauge invariant Hilbert space of a gauge theory is a nontrivial task as it involves solving all the Gauss law constraints. This becomes particularly nontrivial for the gauge groups being non-Abelian. In  Hamiltonian simulation on a classical computer, this results in an exponential rise in computational complexities \cite{PhysRevD.104.074505}. In terms of analog quantum simulation, imposing the Gauss law constraint is an additional burden in the simulation protocol \cite{Zohar:2012xf,Zohar:2013zla,Zohar:2014qma} as the local constraints are not necessarily manifested in the simulating quantum mechanical Hamiltonian. Besides gauge invariance, a theory can have several global symmetries respected by the Hamiltonian, and the dynamics remain confined in each super-selection sectors \cite{klco2018quantum}.  Restricting the dynamics in the physical Hilbert spaceas well as in a particular superselection sectors is essential, yet a challenging task in the NISQ era quantum simulations due to the erroneous quantum hardwares \cite{klco2018quantum, ciavarella2021trailhead, stryker2019oracles, raychowdhury2020solving}.

This has sparked a sharp interest in the past years to find out symmetry protection protocols in a quantum simulation \cite{tran2021faster, halimeh2021gauge, Halimeh_2022, halimeh2020reliability} for simulating a gauge theory and experimental observation of the same \cite{Yang:2020yer}. There have been proposal of studying breaking of gauge-invariance in abelian gauge theories \cite{halimeh2020reliability}. Halimeh et. al studied $\mathbb{Z}_2$ and U(1) gauge theory, whose theoretical and numerical modelling was based on an experimental realization of the same theory \cite{Schweizer2019}. Furthermore, a non-Abelian study of gauge breaking was also proposed \cite{Halimeh_2022} that involves a single body protection protocol. 
In this work, we demonstrate that the complete symmetry protection in the dynamics of $1+1$ dimensional non-Abelian gauge theories can indeed be achieved using a comparatively less involved, namely a global protection protocol.  

The original Kogut-Susskind Hamiltonian for SU(2) gauge theory has been recently mapped to a novel loop-string-hadron (LSH) framework \cite{Raychowdhury:2019iki} without losing any generality. The corresponding Hilbert space is spanned by a set of locally defined LSH basis vectors characterized by manifestly SU(2) invariant local quantum numbers. The Hamiltonian contains both diagonal and ladder operators acting locally on the LSH states. The LSH basis, being gauge-invariant as well as one-sparse, significantly reduces the computational cost in Hamiltonian simulation \cite{PhysRevD.104.074505} and has been found to act as a suitable framework for quantum simulation using both analog \cite{dasgupta2022cold} and digital schemes \cite{raychowdhury2020solving} in the recent past. Albeit being local and SU(2) invariant, the notion of nonlocality of the physical observables of a gauge theory, such as loops or strings,  is still contained in the LSH framework in terms of a set of 'on-link' constraints that is referred as Abelian Gauss law (AGL) constraints in the literature. The physical Hilbert space of the SU(2) gauge theory is thus constructed by local (on-site) LSH states weaved by local (on-link) AGL constraits. 

The Hamiltonian simulation, based on the LSH framework, hence requires imposing local AGL constraints in order to confine dynamics within the physical Hilbert space \cite{dasgupta2022cold, raychowdhury2020solving}.
Following the work of Halimeh et.al \cite{halimeh2020reliability}, we propose a AGL protection term for the LSH Hamiltonian, and explore its implications. Interestingly, the results of this paper demonstrates that, for two dimensional non-Abelian gauge theories, protecting a single abelian global symmetry results in the complete protection of all of the local symmetries generated by AGL. 

The organization of the paper is as follows: The consecutive sections describe the Hamiltonian, Hilbert space and the symmetries of the theory. A symmetry violating term in the Hamiltonian is introduced next and it is discussed how one can protect the symmetry. The protection scheme is validated by numerical evidences of protecting all the symmetries of the theory within this protocol. This work contains explicit results for SU(2), that can be  generalized to SU(3).

\noindent
\section{The Hamiltonian and Hilbert space}
\label{sec:ham}
The LSH Hilbert space is characterized by the local basis states
\begin{align}
    \ket{n_l,n_i,n_0}_{(x)} ~~~\forall x
\end{align}
where the three integer-valued quantum numbers $n_l$, $n_i$, and $n_o$ signify local (on-site) snapshots of electric flux flowing in non-local Wilson loops and strings, as well as ends of a string, and hadrons.  The allowed range of these numbers are
$
    0\leq n_l\leq\infty~~;~~
    n_i,n_o \in \{0,1\}
$, 
designating the loop quantum number $n_l$ to be bosonic and string-quantum numbers $n_i,n_o$ to be fermionic. Presence of a hadron is denoted by non-zero values of both the string quantum numbers in the LSH basis \cite{Raychowdhury:2019iki}. 

The Hamiltonian of the theory in $1+1$ dimension involve the terms:
\begin{eqnarray}
H=H_{\text{electric}}+H_{\text{mass}}+H_{\text{interaction}}
\end{eqnarray}
that correspond to the contributions from total color-electric flux flowing in the lattice, mass of staggered fermions and matter-gauge interactions respectively.  The details of the LSH Hamiltonian can be found in \cite{Raychowdhury:2019iki}.

The LSH basis introduced above is a strong coupling eigen basis of the Hamiltonian, in which the electric and mass terms of the Hamiltonian are diagonal.
The matter gauge interaction term for $1$-d spatial lattice is responsible for interesting dynamics of the  theory. This particular term, being gauge invariant, acting on the strong coupling vacuum (a gauge invariant state) build up the gauge invariant Hilbert space. This term within the LSH framework \cite{Raychowdhury:2019iki} consists of local SU(2) invariant  creation/annihilation operators corresponding to `string-ends' located at  nearest neighbour sites, 
\begin{eqnarray}
\label{Hi}
    \hat{H}_{\text{int.}} &=&x_0 \sum_{x} \hat\eta(x) \left[\sum_{\sigma=\pm} \mathcal{S}_{\text {out }}^{+, \sigma}(x) \mathcal{S}_{\text {in }}^{\sigma,-}(x+1)\right]\hat \eta(x+1)\nonumber  \nonumber \\&& +\text { h.c. }
\end{eqnarray}
Where, $x_0$ is a dimensionless coupling and $\hat\eta(x)$ are diagonal operators in the LSH basis.  The string-end operators $\mathcal{S}_{\text {out }}^{\pm, \sigma}(x)$ and $\mathcal{S}_{\text {in }}^{\sigma,\pm}(x)$ are manifestly SU(2) singlet operators with both bosonic and fermionic field content that denote start/end of a string (out/in). These string-end operators create/annihilate ($\pm$) fermionic field at both its ends and changes the electric flux on both ends of the link connecting neighboring sites in order to preserve local SU(2) invariance.

Albeit being constructed out of local SU(2) singlets, for the LSH interaction Hamiltonian to describe the full dynamics of the theory, must incorporate continuation of the strings across neighboring sites aka AGL. This is guaranteed by the same $\sigma$ index in both the string end terms located at sites $x$ and $x+1$. The index $\sigma=\pm$ denotes creation/annihilation of one unit of electric flux along the link connecting sites $x$ and $x+1$.  As a conseqence of (\ref{Hi}) satisfying the AGL, the eigenvalues of $\hat\eta(x)$ and $\hat\eta(x+1)$ in (\ref{Hi}) are identical while acting on a physical state, that is annihilated by AGL given in (\ref{AGL-def}). 

\section{The symmetries}
\label{sec:GS}
In addition to the local SU(2) gauge symmetry, that is manifested in the SU(2) invariant LSH framework, the LSH Hamiltonian in $1+1$-d contains the following additional symmetries listed below:
\begin{itemize}
    \item {\textbf{Local symmetries imposed by AGL: }} The LSH states $|n_l,n_i.n_o\rangle_x$ at neighboring sites must satisfy the local `on-link' constraint:
    \begin{eqnarray}
    \label{AGL-def}
    n_l+n_o(1-n_i)\Big{|}_{x}= n_l+n_i(1-n_o)\Big{|}_{x+1}
    \end{eqnarray}
    \item{\textbf{Global symmetries: }} In addition to the SU(2) gauge symmetries, the LSH Hamiltonian also admits global SU(2) symmetries \cite{PhysRevD.104.074505} and hence a global LSH state is characterized by the global quantum numbers that correspond to the total angular momentum and the magnetic quantum number for global SU(2). In the LSH framework, these two global quantum numbers are mapped to the following Global quantities:
    \begin{enumerate}
        \item Total fermionic occupation number:
\begin{eqnarray}
\label{Q}
   Q = \sum_{x=0}^{N-1}\big[{n}_i(x)+{n}_o(x)\big]  
\end{eqnarray}
For a $N$-site lattice, the value of $Q$ can be any integer between $[0,2N]$. 
\item The imbalance between incoming and outgoing strings: relates to the boundary fluxes
\begin{equation}
\label{q}
    {q} = \sum_{x=0}^{N-1}\big[{n}_0(x)-{n}_i(x)\big]
\end{equation}
For a particular $Q$ value, ${q}$ can take any value from $-Q$ to $+Q$ and defines different disconnected sectors of the larger gauge-invariant LSH Hilbert space.
\end{enumerate}
The LSH Hamiltonian obeys both the $Q,q$ symmetries and in turn results in a block diagonalized structure. The dynamics of the theory remain confined within each block, enabling computational benefit \cite{PhysRevD.104.074505}.  
\item{\textbf{Charge conjugation symmetry: }} The particle anti-particle symmetry of the theory identifies $(Q,q)$ sector of the Hamiltonian to the $(Q,-q)$ sector. 
 \end{itemize}

\subsection{An observation: relating local and global symmetries}
As previously mentioned, the AGL invariance of the LSH Hamiltonian is manifested by the same $\sigma$ indices for both the ends of the string operator in (\ref{Hi}). At this point, taking a deeper look into the LSH Hamiltonian, specifically at the off-diagonal elements of the Hamiltonian written in terms of the normalized ladder operators (defined in \cite{Raychowdhury:2019iki}) is necessary. The string end operators in the interaction Hamiltonian (\ref{Hi}) written in terms of the normalized ladder operators in the LSH basis, takes the form \cite{Raychowdhury:2019iki}:
\begin{eqnarray}
\label{ss1}
    \hat{S}_{out}(x)^{++}\hat{S}_{in}^{+-}(x+1) &\approx \hat{\chi}_{o}^{\dagger}(x)\hat{\chi}_{o}(x+1) \label{s1}\\
    \hat{S}_{out}^{+-}(x)\hat{S}_{in}^{--}(x+1) &\approx \hat{\chi}_{i}^{\dagger}(x)\hat{\chi}_{i}(x+1) \label{s2}\label{ss2}
\end{eqnarray}
where, the action of the normalized ladder operators on the local LSH quantum numbers are realized as\footnote {In (\ref{ss1},\ref{ss2}), the left and right hand sides are related by $\approx$, as the full equality includes suitable ladder operators in the loop quantum number $n_l$ as well as normalization factors \cite{Raychowdhury:2019iki}. The approximated form is sufficient for understanding the symmetry protection protocol developed in this work. However, the numerical calculations use the exact expression given in \cite{Raychowdhury:2019iki}.} :
    \begin{eqnarray}
    \chi^\dagger_{i}|n_l,n_i,n_o\rangle &=& (1-n_i)|n_l,n_i+1,n_o\rangle  \\
    \chi_{i}|n_l,n_i,n_o\rangle &=& n_i|n_l,n_i-1,n_o\rangle  \\
    \chi^\dagger_{o}|n_l,n_i,n_o\rangle &=& (1-n_o)|n_l,n_i,n_o+1\rangle  \\
    \chi_{o}|n_l,n_i,n_o\rangle &=& n_o|n_l,n_i,n_o-1\rangle 
    \end{eqnarray}
    
From the above set of equation, one can readily suggests that the pairwise presence of the the outgoing string operator at site $x$ and the incoming string operator at site $x+1$, with the same $\sigma$ index, i.e $\sigma=+$ for (\ref{s1}) and $\sigma=-$ for (\ref{s2}) in effect preserves the total number of $n_o$ excitation and $n_i$ excitation respectively on a pair of neighboring sites. The entire Hamiltonian, with interaction terms present for each and every neighboring site, that satisfy the AGL, in turn preserves the global quantum numbers $\sum_x n_i(x)$ and $\sum_x n_o(x)$ for the lattice. 

\section{Violating the symmetries}
\label{sec:violate}
\noindent
Even with guaranteed protection of the non-Abelian gauge symmetries while working with the LSH framework for the same, several possible sources of error that is present for both analog and digital simulation schemes would most likely break the Abelian gauge invariance of the LSH dynamics and hence would take the dynamics away from the physical Hilbert space. As an example, the analog simulation scheme of simulating LSH dynamics (as in \cite{dasgupta2022cold}), where the two fermionic LSH degrees of freedom are mapped to  up/down spins of the neutral atoms respectively may experience spin-flip error , that would result in violation of AGL in the dynamics as well as would couple multiple super-selection sectors of the theory.  For a digital simulation scheme, bit-flip errors would also cause a similar AGL violation. Such a consideration for a digital simulation has previously led to the construction of Gauss law oracles \cite{stryker2019oracles, raychowdhury2020solving} to check for such error. We model this particular error by adding an extra term in the LSH Hamiltonian that would always violate the AGL and drag the dynamics away from the physical Hilbert space. 

A term, that would not satisfy AGL can be the same interaction Hamiltonian (\ref{Hi}), but not necessarily with the same $\sigma$ index as given below:
\begin{eqnarray}
\centering
\label{H'i}
    \hat{H'}_{\text{int.}} &=& x_0\sum_{x} \hat\eta(x) \left[\sum_{\sigma,\sigma'=\pm} \mathcal{S}_{\text {out }}^{+, \sigma}(x) \mathcal{S}_{\text {in }}^{\sigma',-}(x+1)\right]\hat \eta(x+1)\nonumber  \nonumber \\ +\text{ h.c. }
\end{eqnarray}
Note that, violation of AGL would imply the eigenvalues of the diagonal operators $\hat\eta(x)$ and $\hat\eta(x+1)$ can be different, unlike the case in (\ref{Hi}). 
This particular form of the interaction Hamiltonian contains the desired AGL preserving interaction (\ref{Hi}) as well as terms that violate AGL with equal weight, but can be tuned separately to model individual cases as par the simulation schemes. 

 \begin{figure*}[!htb]
    \centering \includegraphics[width=2\columnwidth]{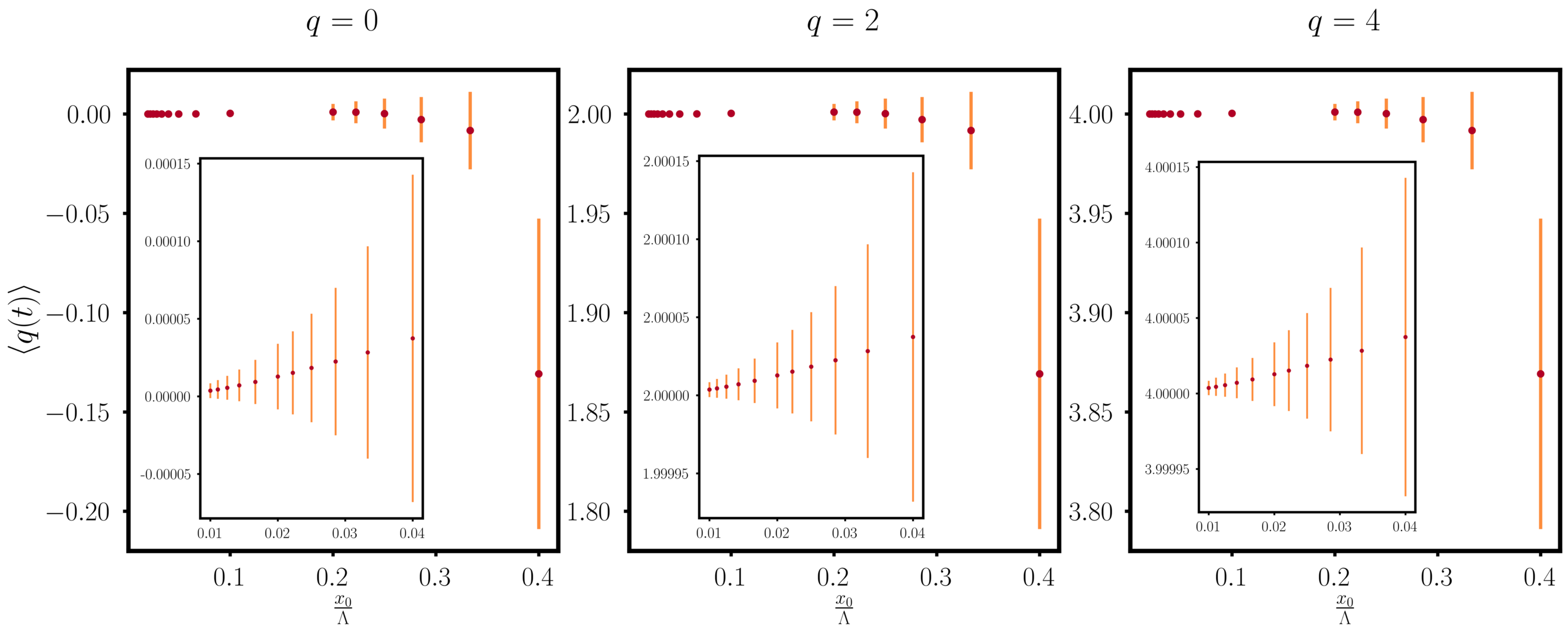}
    \caption{Time average (for long time) of the global quantum number $q=\sum_x(n_o(x)-n_i(x))$   is plotted against the ratio of  $x_0$, the dimensionless couplings that come with the symmetry violating interaction,  and $\Lambda$ the coupling with the symmetry protection term. The error bar denotes the deviation of the observed value from the mean time averaged value throughout the dynamics. With stronger protection strength, i.e $\frac{x_0}{\Lambda}\rightarrow 0$  the global symmetry is better protected as manifested in the zoomed in region shown in the inset. The plots illustrate that the dynamics is successfully projected within each global symmetry sector $q=0,2,4$ with each protection term $H^{q}_{\textrm{protect}}$.}
    \label{plot1}
\end{figure*}

\section{Protecting the symmetry}
\label{sec:protect}
As previously stated, the local constraints in the LSH framework, given by the AGL's in turn result in protection of global quantum numbers $\sum_x n_i(x)$ and $\sum_x n_o(x)$ for the lattice, the violation of AGL should directly imply not conserving the same. This statement can be mathematically demonstrated by writing the AGL violating interaction term introduced in (\ref{H'i}) in terms of the normalized ladder operators as in (\ref{ss1},\ref{ss2}) as follows:
\begin{eqnarray}
\label{ss3}
    \hat{S}_{out}(x)^{++}\hat{S}_{in}^{--}(x+1) &\approx \hat{\chi}_{o}^{\dagger}(x)\hat{\chi}_{i}(x+1) \label{b1}\\
    \hat{S}_{out}^{+-}(x)\hat{S}_{in}^{+-}(x+1) &\approx \hat{\chi}_{i}^{\dagger}(x)\hat{\chi}_{o}(x+1) \label{b2}\label{ss4}
\end{eqnarray}
From (\ref{ss3}) and (\ref{ss4}), it is evident that the quantities $\sum_x n_i(x)$ and $\sum_x n_o(x)$ are not being conserved in presence of the interaction Hamiltonian (\ref{H'i}).  We further note that, the global quantities $\sum_x n_i(x)$ and $\sum_x n_o(x)$ are nothing but the linear combination of the previously introduced global SU(2) total angular momentum and its third component given by $Q,q$ in (\ref{Q},\ref{q}). As the AGL violating interaction term (\ref{H'i}) still conserved the global quantity $Q$ as per (\ref{ss1},\ref{ss2},\ref{ss3},\ref{ss4}), we relate that the violation of all of the local constraints is practically equivalent to violation of the global $q$ symmetry.

In this regard, we propose that  the dynamics of an erroneous Hamiltonian given by:
$$H'=H_{\textrm{electric}}+H_{\textrm{mass}}+H'_{\textrm{interaction}}$$
can be made to remain confined in the physical subspace consisting of Wilson loops, strings and hadrons by adding the following global protection term to the above Hamiltonian:
\begin{eqnarray}
\label{Hpr}
H^{q}_{\textrm{protect}}= \Lambda\left[q-\sum_{x=0}^{N-1}\left( \hat{n}_i(x)-\hat{n}_o(x)\right)\right],
\end{eqnarray}
where, $q$ can be any integer in $[-Q,Q]$ that projects the dynamics of the erroneous Hamiltonian in a particular $(Q,q)$ sector. The bonus of projecting the dynamics to any of the particular $Q,q$ sector is the automatic validation of the AGL constraints throughout the lattice as demonstrated by the numerical results in the next section.
  \begin{figure*}[!htb]
    \centering \includegraphics[width=2\columnwidth]{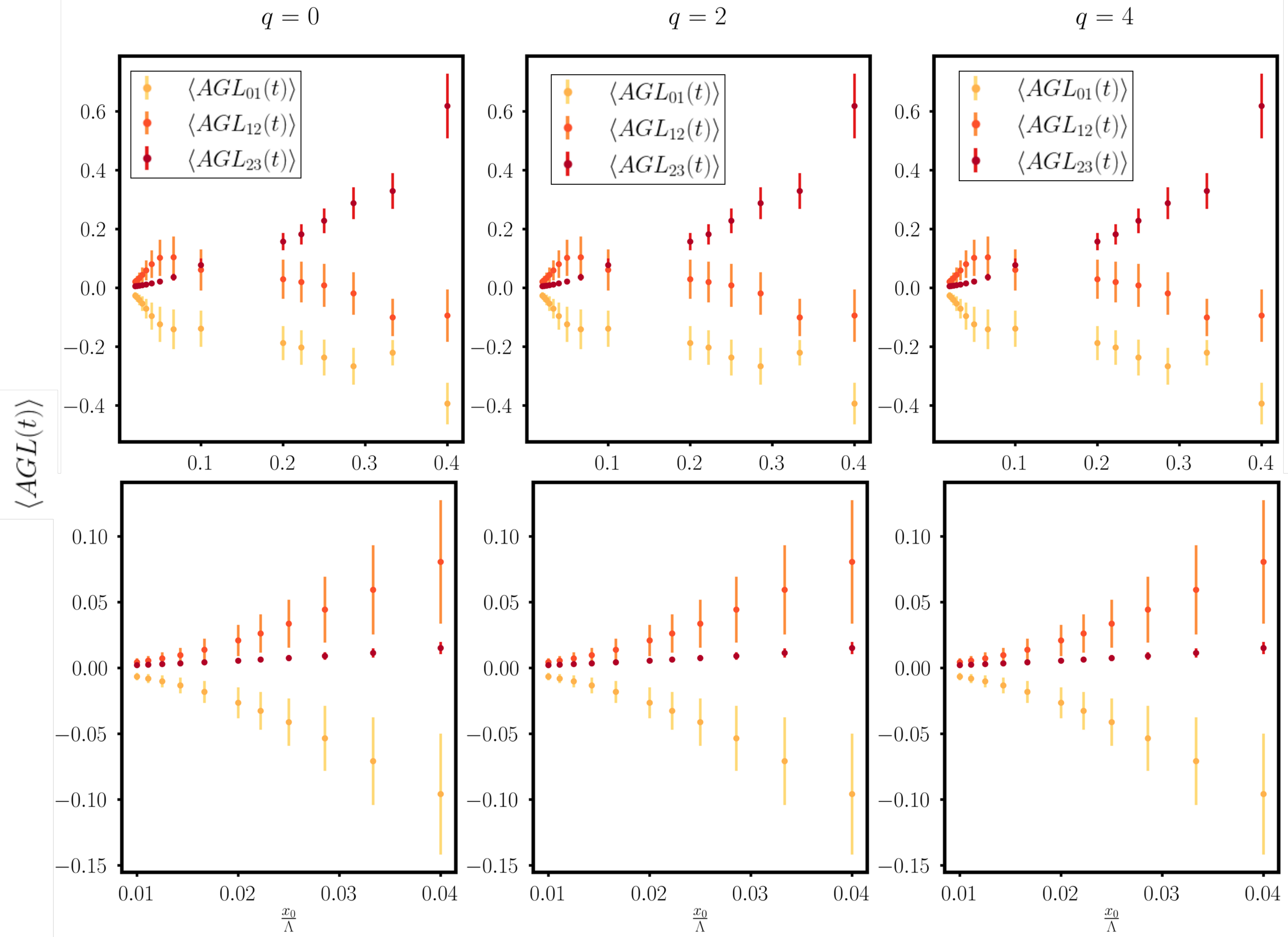}
    \caption{The time averaged value of the AGL observables $AGL_{x,x+1}$ on each link of the 4 site lattice, while time evolved with the Hamiltonian that contains local AGL violating interaction terms with strength $x_0$ and global symmetry protection term with strength $\Lambda$, is plotted against $x_0/\Lambda$. For any $q$-sector of the protected global symmetry, the Abelian Gauss laws on all of the links are perfectly protected with increasing protection strength. The error bars denote the fluctuation of the obserbavles from its mean value during the evolution. The second row of plots demonstrate systematically better protection (including systematic decrease in fluctuations as well) of all of the Abelian Gauss laws as $x_0/\Lambda$ is decreased upto $1/100$ for a 4-site lattice.}
    \label{plot2}
\end{figure*}

\section{Results}
\label{sec:result}
The time evolution of a physical state under the erroneous Hamiltonian $H'$ along with the protection term, i.e.  \begin{eqnarray}
\tilde H= H'+H^{q}_{\textrm{protect}}
\end{eqnarray} is studied for different ratios of the dimensionless couplings $x_0/\Lambda$, where, $x_0$ is the dimensionless coupling coefficient with the interaction Hamiltonian in (\ref{H'i}) and $\Lambda$ is the dimensionless protection parameter in (\ref{Hpr}). 

Time evolution of the following  two observable is studied:
\begin{enumerate}
\item Global observable:
\begin{align*}
    q(t) = \bra{\Psi(t)}\sum_{x=0}^{N-1}\Big[\hat{n}_i(x)-\hat{n}_o(x)\Big]\ket{\Psi(t)}
\end{align*}
\item Local observable, that measures AGL for each link connecting sites $x$ and $x+1$:
\begin{align}
    AGL_{x,x+1}(t) = \bra{\psi(t)}\left[\mathcal{N}_L(x)-\mathcal{N}_R(x+1)\right] \ket{\psi(t)}
\end{align}
where, 
\begin{eqnarray}
\centering
    \mathcal{N}_L(x)\ket{n_l,n_i,n_o}_x  &=& [n_l+ n_o(1-n_i)]_x\ket{n_l,n_i,n_o}_x \nonumber \\
\mathcal{N}_R(x+1)\ket{n_l,n_i,n_o}_{x+1}  &=& \big[n_l+ n_i(1-n_o)\big]_{x+1}\ket{n_l,n_i,n_o}_{x+1} \nonumber 
\end{eqnarray}
\end{enumerate}
by exact diagonalization technique for a 4-site lattice, and the time evolution is studied using QuSpin \cite{SciPostPhys.2.1.003,10.21468/SciPostPhys.7.2.020} for sufficiently large time $T$ at small intervals of $0.0001T$. The results presented are for open boundary condition, with zero incoming flux at the $0^{th}$ lattice site.

The numerical result is summarized below:
\begin{itemize}
    \item Figure (\ref{plot1}) demonstrates that $\langle q(t)\rangle$, time average of the global quantum number $q$ converges to the value $\langle q(t)\rangle=q$  by using protection Hamiltonian $H^{q}_{\textrm{protect}}$ with increasing protection strength, (i.e. decreasing $x_0/\Lambda$) for $ q=0,2,4$.
     \item Figure (\ref{plot2}) demonstrates that $\langle AGL_{x,x+1}(t)\rangle$, time average of the AGL quantum number  for each link $(x,x+1)$ converges to the value $0$  by using protection Hamiltonian $H^{q}_{\textrm{protect}}$ with increasing protection strength, (i.e. decreasing $x_0/\Lambda$) for all the values of $q=0,2,4$.
\end{itemize}

\section{Conclusion and outlook}
\label{sec:conclusion}
This article presents a novel idea of simulating the physical dynamics of a non-Abelian gauge theory in two dimensions described by a local Hamiltonian without imposing any local constraints. This was only made possible within the LSH framework, where the complete Abelianization of SU(2) gauge symmetries have been performed without introducing any non-local interaction. For $1+1$ dimensional case, the whole set of local Abelian symmetries has been made captured by a single global symmetry that is manifested in the LSH Hamiltonian construction as well.  The whole study of protecting the symmetry  is based on the idea of imposing a constraint in Hamiltonian dynamics with a large value of Lagrange multiplier. The current study on small lattice can be easily generalized to to use state-of-the-art Tensor network techniques to probe for larger system\cite{Carmen_Ba_uls_2020}. It is well know that tensor networks can approximate ground states for 1D lattice theories.\cite{Or_s_2014}. Encoding global symmetries within the tensor network formulation is possible, and since gauge breaking in the LSH formulation is a global symmetry violation, it is possible to use the tensor network formulation to study these dynamics. It might even be possible to explore larger lattice sizes so that one can study the scaling effect of the protection strength.

This particular study removes a vast set of the difficulties (regarding imposing all the constraints) in quantum simulating non-Abelian gauge theories in lower dimension within the scope of NISQ ers devices and also for tensor network calculations as the framework becomes free of local symmetries and yet the interactions remain local. A similar global protection scheme for a SU(3) gauge theory \cite{Anishetty:2009ai, Anishetty:2009nh} is also possible to construct, provided there exists a  LSH framework for SU(3) as well. Another important aspect of this study is to provide clear insight into the entanglement structure for a $1+1$-dimensional non-Abelian gauge theory. Works are in progress in these directions and will be reported shortly. 

\noindent
\section*{Acknowledgments}
We would like to thank Aniruddha Bapat, Zohreh Davoudi, Rudranil Basu and Raka Dasgupta for useful discussions. IR is partly supported by the Research Initiation Grant (reference no. BPGC/RIG/2021-22/10-2021/01) and OPERA award (reference no. FR/SCM/11-Dec-2020/PHY) at BITS-Pilani.
\noindent

\bibliography{bibi}
\end{document}